\documentclass[a4paper,11pt]{article}

\usepackage{jcappub} 

\usepackage[T1]{fontenc} 

\title{\boldmath Photon and Neutrino Spectra of Time-Dependent
Photospheric Models of Gamma-Ray Bursts}

\author[a]{K. Asano}
\author[b,1]{and P. M\'esz\'aros\note{Corresponding author.}
}
\affiliation[a]{Institute for Cosmic Ray Research, The University of Tokyo,
5-1-5 Kashiwanoha, Kashiwa, Chiba 277-8582, Japan}
\affiliation[b]{Department of Astronomy \& Astrophysics; Department of Physics;
Center for Particle \& Gravitational Astrophysics;
Pennsylvania State University, University Park, PA 16802}

\emailAdd{asanok@icrr.u-tokyo.ac.jp}
\emailAdd{nnp@astro.psu.edu}

\abstract{Thermal photons from the photosphere may be the primary source
of the observed prompt emission of gamma-ray bursts (GRBs). In order 
to produce the observed non-thermal spectra, some kind of dissipation 
mechanism near  the photosphere is required.
In this paper we numerically simulate the evolution of the photon 
spectrum in a relativistically expanding shell with a time-dependent 
numerical code. We consider two basic models. One is a leptonic model, 
where a dissipation mechanism heats the thermal electrons maintaining 
their high temperature. The other model involves a cascade process induced 
by $pp$($pn$)-collisions which produce high-energy
electrons, modify the thermal 
spectrum, and emit neutrinos.  The qualitative properties of the photon 
spectra are mainly determined by the optical depth at which the dissipation
mechanism sets in. Too large optical depths lead to a broad and curved 
spectrum contradicting the observations, while for optical depths smaller
than unity the spectral hardness becomes softer than observed.
A significant shift of the spectral peak energy to higher energies
due to a large energy injection can lead to an overly broad spectral shape.
We show ideal parameter ranges for which these models are able to reproduce the 
observed spectra.  For the $pn$-collision model, the neutrino fluence in the 
$10$--$100$ GeV range is well above the atmospheric neutrino fluence, but its 
detection is challenging for presently available detectors.}

\begin{document}
\maketitle
\flushbottom

\section{Introduction}
\label{sec:intro}

The typical spectrum of the prompt emissions
of gamma-ray bursts (GRBs) has a peak
at the peak energy $\varepsilon_{\rm p}$ of $0.1-1$ MeV
in $\varepsilon f(\varepsilon)$-plot.
To express the spectral shape,
the Band function \cite{ban93} has been frequently used.
This function smoothly join two power-law parts
for high-energy (photon index $\beta$) and low-energy
(index $\alpha$) regions at $\varepsilon_{\rm p}$.
However, the origin of this function is not understood yet.
One of the most famous models is synchrotron emission
from shock accelerated electrons.
In this model, the peak energy $\varepsilon_{\rm p}$
is attributed to the typical photon energy
emitted from the lowest-energy electrons at injection.
However, the lowest energy $\varepsilon_{\rm e,min} \gg m_{\rm e} c^2$
is phenomenologically assumed,
and we have no definite idea for
the physical mechanism to determine $\varepsilon_{\rm e,min}$.
Nevertheless, in most cases, the peak energy
does not show drastic variation and seems relatively steady in a GRB,
even though its lightcurve shows multiple pulses \cite{for95}.
Another problem in the synchrotron model is the low-energy spectral index.
In this model injected electrons are promptly cooled,
and the distribution of the cooled electrons should be a power-law
of $n_{\rm e}(\varepsilon_{\rm e}) \propto \varepsilon_{\rm e}^{-2}$.
This distribution predicts $\alpha=-1.5$,
while the typical fitted value is $-1.0$ \cite{pre00}.
To make matters worse, some GRBs show harder spectra
than the ``death-line'' index $-2/3$, which is not explained
by usual synchrotron emission \cite{pre00}.

One of the most promising alternative models
is the photosphere models,
in which the thermal photons emitted from
the photosphere construct the dominant spectral component.
The thermal radiation is naturally expected by the original
fireball model \cite{pac86,pir93,mes93}, in which
radiation pressure accelerates the outflow to ultra relativistic
velocity (the Lorentz factor $\Gamma=100$--$1000$).
In the simplest case (without pair loading etc.), the photosphere radius is
\begin{eqnarray}
R_{\rm ph}=\frac{L_{\rm iso} \sigma_{\rm T}}
{4 \pi \eta^3 m_{\rm p} c^3}
\simeq 5.4 \times 10^{11} \left( \frac{\eta}{600} \right)^{-3}
\left( \frac{L_{\rm iso}}{10^{53}~\mbox{erg}~\mbox{s}^{-1}} \right)~\mbox{cm},
\end{eqnarray}
where $L_{\rm iso}$ is the isotropic-equivalent total luminosity
and $\eta$ is the baryon loading parameter,
which corresponds to the final Lorentz factor after the bulk acceleration
is saturated.
The magnetic field and electron--positron pairs etc.
can contribute some an uncertainty in the photosphere radius.
The wide-band observations with {\it Fermi} have
explored the possibility of the photosphere model.
Guiriec et al. (2011) \cite{gui11} claim the detection of a thermal component
in the spectrum of GRB 100724B, though another Band component
is dominant in the entire spectrum.
The very narrow spectral component in GRB 090902B
is the most impressive example that is consistent with
the thermal spectrum \cite{ryd11,pee12}.

However, such examples are exceptional.
In most cases, the prompt spectra distribute in a wide energy range.
So efficient energy dissipation of the outflow
is required to modify the thermal spectrum.
Such dissipative photosphere models have been proposed
by several authors \cite[e.g.][]{mes00,gia06,pee06,iok07,laz10}
with some kind of energy dissipation,
such as shock, plasma instability, or magnetic reconnection.
One of encouraging models, Beloborodov (2010) \cite{bel10} consider
the cascade process induced by $pn$-collision as the dissipation mechanism.
The electron--positron pairs injected via the cascade up-scatter
the thermal photons.
The model shows a similar spectrum to the observed Band function
\cite[see also][]{vur11}.

In this paper we investigate universality of the Band function
in the dissipative photosphere models.
Our one-zone time-dependent simulation is based on the shell picture.
While the radiative transfer is solved in a steady flow
in \cite{bel10} and \cite{gia07} etc.
\cite[see also][for similar simulations
of the AGN photosphere]{asa07,asa09},
a relativistically expanding shell has been frequently
considered as an emission region.
A more realistic description would probably lie between these 
steady-flow and shell picture extremes. 
Thus, here we adopt a different type of simulation,
based on a one-zone time-dependent approach,
as a first step towards a more meaningful
description of a  dissipative photosphere.
A precedent for such time-dependent simulations
is \cite{pee06}, where the spectral evolution,
given the dissipative radius,
is followed only within the dynamical timescale.
Here, we simulate the spectral evolution
from the onset of the dissipation until
photons escape from the emission region.

In \S \ref{sec:model} we provide an outline of the model
and code of the numerical simulation.
Here we investigate both models with gradual dissipation and
models with a sudden onset of the energy dissipation, 
where the spectrum is formed by either
(1) photon scattering by thermal electrons heated by the dissipation
or (2) $pn$($pp$)-collision model similar to \cite{bel10}.
The results of the photon spectrum for those models
are shown in \S \ref{sec:lep} and \S \ref{sec:had}.
The hadronic interaction leads to neutrino emission.
The neutrino spectrum is also shown and discuss the detectability.
We also show lightcurves obtained from our results in \S \ref{sec:LC}.
The summary and discussion are in \S \ref{sec:summary}.
To provide spectra and lightcurves for an observer,
we adopt the cosmological redshift $z=2$ throughout this paper
with the cosmological parameters $H_0=70~\mbox{km}~\mbox{s}^{-1}~\mbox{Mpc}^{-1}$,
$\Omega=0.3$, and $\Lambda=0.7$.

\section{Model and Method}
\label{sec:model}

We use the same numerical code in \cite{asa11,asa12}
that can follow the temporal evolutions of energy distributions
of photons, electrons/positrons, protons, neutrons, pions,
muons, and neutrinos in a relativistically expanding shell.
With this code we simulate single pulse emitted from the shell.
The simulation starts at a radius $R=R_0$, where some kind of dissipation mechanism
onsets. Initially, thermal photons with a diluted Planck spectrum
and thermal electrons
are confined in a relativistically expanding shell
with the bulk Lorentz factor $\Gamma$.
The geometry of the shell is spherically symmetric,
but we only consider contribution of photons escaping
from the surfaces within a jet opening angle $\theta_{\rm j}$.
The evolutions of the particle-energy distributions
are calculated taking into account
the interactions between photons and electrons,
synchrotron emission, and photon escape.
For hadronic models, we also include the effects
of $pp$ and $p \gamma$ collisions, and secondary particles
as explained in \cite{asa12}.
Since our code is one-zone, all particles are assumed to
be homogeneous and isotropic in the shell rest frame
in spite of the photon escape from the shell surfaces.
Hereafter, we denote amounts in the shell rest frame with
primed characters.
The initial electron density is parameterized by the optical depth $\tau_0$ as
\begin{eqnarray}
n'_{\rm e}(R_0)=\tau_0 \frac{\Gamma}{\sigma_{\rm T} R_0}.
\end{eqnarray}
Photons and neutrinos escape from both posterior and anterior surfaces
of the shell.
Our code does not calculate the radiative transfer in the shell.
To include the effect of scattering on photon escape,
we introduce the effective photon velocity as
\begin{eqnarray}
c'_{\rm eff} \equiv \min \left(
c,c/\tau \right),
\end{eqnarray}
where $\tau=n'_{\rm e} \sigma_{\rm T} R/\Gamma$.
For a time step $\Delta t'$, we roughly approximate the escape fraction of photons
as $c'_{\rm eff} \Delta t'/(2 W')$,
where $W'$ is the shell width.
The electron density evolves with shell expansion and
injection of secondary particles.
Counting the photons escaping from the shell for each timestep,
we follow the spectral evolution for an observer
with the curvature effect of the relativistically expanding shell
\cite[for details, see][]{asa11}.

We fix the model parameters as follows:
the initial radius where the dissipation starts
$R_0=10^{11}$ cm,
bulk Lorentz factor of the shell $\Gamma=600$,
jet opening angle $\theta_{\rm j}=10/\Gamma$,
initial photon temperature $T'_{\rm th}=0.2$ keV,
and average luminosity of the thermal photons
$L_{\rm th}=4 \pi c R_0^2 \Gamma^2 U'_{\rm th}=10^{52} \mbox{erg}~\mbox{s}^{-1}$,
which determines the photon energy density $U'_{\rm th}$ at $R=R_0$
with the above parameters.

The parameters we adjust below are the initial optical depth $\tau_0$,
shell width $W=\Gamma W'$, magnetic field $B'$, electron heating rate
due to dissipations, and hadronic contribution.

\section{Leptonic Models: Spectral Shapes}
\label{sec:lep}

Our code can treat the photon/particle production
and corresponding cooling of particles.
However, in this paper, we consider not only cooling
but also heating due to some unknown energy-dissipation.
In the dissipative-photosphere models, the required dissipation energy should be
at least comparable to the initial thermal photon energy.
If the energy injection around the photosphere is impulsive,
the average energy per electron largely exceeds the rest mass energy
as with the usual internal shock model.
In this case, as shown in \cite{pee06}, inverse Compton (IC)
emission appears as a different spectral component
rather than modification of the Planck spectrum.
Therefore, to transfer enough energy to photons,
electrons should be heated continuously maintaining their
non-relativistic temperature.

The radial profile of the electron-heating rate is not well known,
and models differ widely. For example, in the model of \cite{gia06} 
\cite[see also][]{gia12}, the electrons are continuously heated over a 
wide range of distances at a rate $\dot{E} \propto R^{-1}$ starting from a 
sub-photospheric radius, and the electron temperature increases with radius 
as $T'_{\rm e} \propto R$--$R^{5/3}$.  In this model magnetic reconnection 
is supposed to be the heating source. However, the specific mechanism of 
magnetic reconnection in GRB outflows is not well understood, and it is 
unclear whether the dissipation is as efficient and long-lasting as this 
model assumes. A different model of magnetized outflows, based on a  
``reconnection switch'', is proposed by McKinney \& Uzdensky (2012)
\cite{mck12} to heat the electrons.
In yet other models, based on shock formation \cite{pee06,iok07},
a sudden onset of the energy dissipation at a certain radius is assumed.
In this paper we consider these various cases, 
assuming that the dissipation sets in at some radius $R_0$ somewhere below
the photopshere, where the electrons are heated promptly, resulting in a 
temperature jump $T'_{\rm e}>T'_{\rm th}$.
For the gradual dissipation model, we take $R_0$ far enough below
the photosphere that the exact value of $R_0$ is not important and
a behavior consistent with gradual dissipation is approximated.
The high photon density postulated in the photosphere models implies
a very short electron-cooling timescale due to Compton scattering.
To save computational cost, we do not solve the energy equation for 
the thermal electrons.  Instead, we assume
a power-law evolution of the electron temperature of the form
\begin{eqnarray}
T'_{\rm e}=T_0 (R/R_0)^{-s},
\label{eqTev}
\end{eqnarray}
which practically provides the evolution of the heating rate.
The initial temperature $T_0$ and the index $s$ may be determined
by details of a specific dissipation mechanism, but 
here we do not specify $s$, in order to investigate the generic behavior.
Those parameters are treated as free parameters.

\begin{figure}[htb!]
\centering
\includegraphics[width=.75\textwidth]{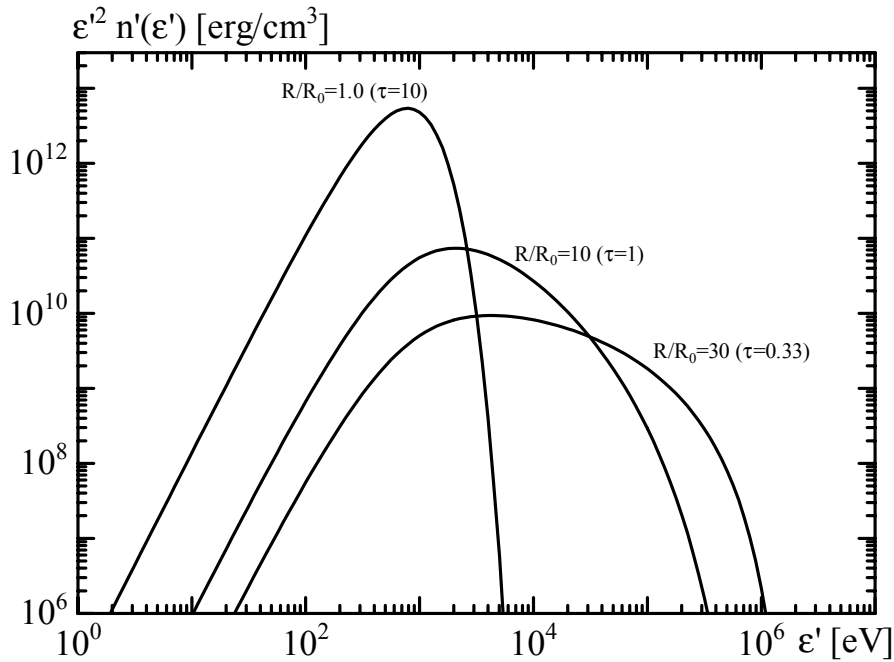}
\caption{Gradual energy dissipation model: evolution of the energy 
distribution of photons in the shell rest frame.
The electron temperature evolution is assumed to be
$T'_{\rm e}=5$ keV $(R/R_0)$, and
the initial optical depth $\tau_0$ is 10.
\label{fig:1}}
\end{figure}

In the standard internal shock model, the shell width
is conventionally assumed as $W'=R_0/\Gamma$,
which is equivalent to the causal distance within the dynamical timescale.
The pulse-timescale $W/c$, estimated from the above width
and photosphere radius ($\lesssim 10^{11}$ cm), would be shorter
than ms, while the observed timescales are typically $\sim 0.1$ s
\cite{nor96}.
In order to reconcile this with the pulse-timescale in photosphere models,
a long acting engine with a pulse duration of $\sim 0.1$ s can be plausibly assumed.
This may justify the steady-flow assumption in the calculations
of \cite{gia07} or \cite{bel10}.
To simulate this situation by our time-dependent code,
we adopt a thick shell of width $W=W'/\Gamma=10^3 R_0/\Gamma^2
\sim 2.8 \times 10^8$ cm in this section.
This width and a luminosity of the thermal photons
$L_{\rm th}=10^{52} \mbox{erg}~\mbox{s}^{-1}$ imply
an initial photon energy in the shell of
$E_{\rm th}=9.3 \times 10^{49}$ erg.
This shell is still thin compared to $R_0$ so that
the homogeneous density in the shell is a rough but acceptable approximation.
If the dissipation is caused by plasma instabilities
or interaction of two flows, the effects would spread
throughout the volume of the shell.
This homogeneous shell assumption may be ad-hoc, because the entire 
volume of the photosphere is not causally connected.  However, we consider 
that the average evolution of the photon distribution in the shell 
is not severely affected by this homogeneous assumption.
The thick shell enlarges the effective radius where
most of all photons escape to $\sim 10^3 R_0$
(we follow the evolution of the shell
until $3 \times 10^4 R_0$).
So photons can interact with the electrons in the shell
longer than the dynamical timescale at the photosphere,
which may resemble the steady-flow approximation.
Actually, as shown in figure \ref{fig:1},
our calculation reconfirms the results
in the preceding steady-flow simulations.
The initial temperature $5$ keV at $\tau=10$ expresses
the situation just after the decoupling of the temperatures
of the electrons and photons.
Then, the electron temperature gradually increases with radius
as in the models of \cite{gia06} or \cite{gia12}.
Such long-lasting energy dissipation, as shown by the above papers,
successfully leads to a ``Band-like spectrum''.

\begin{figure}[htb!]
\centering
\includegraphics[width=.75\textwidth]{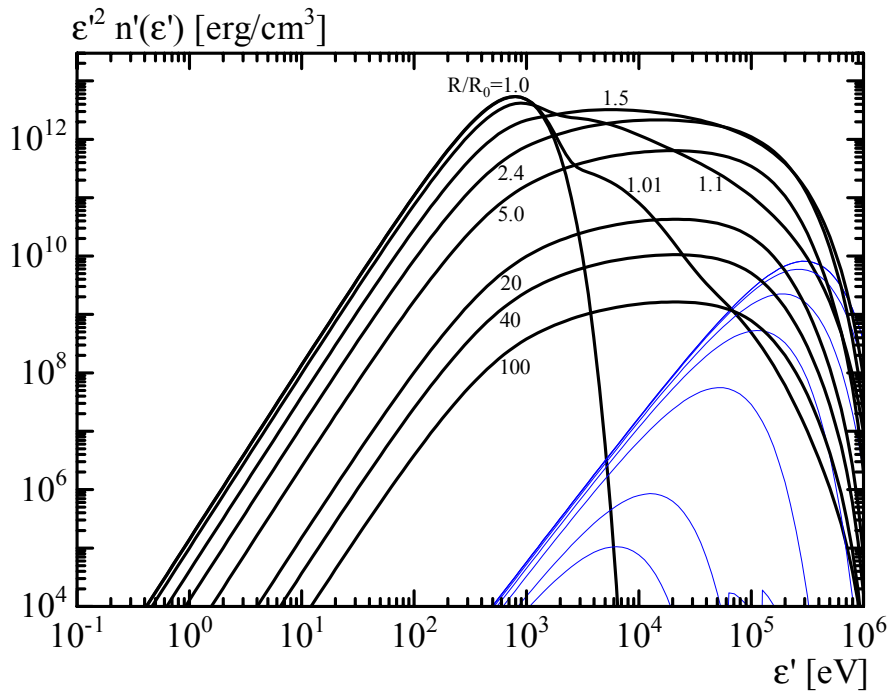}
\caption{Leptonic dissipative-photosphere models: evolution of the energy 
distribution of photons (thick)
and electrons (thin) in the shell rest frame.
The electron temperature evolution is assumed to be
$T'_{\rm e}=100$ keV $(R/R_0)^{-1}$, and
the initial optical depth $\tau_0$ is 5.
\label{fig:2}}
\end{figure}

\begin{figure}[htb!]
\centering
\includegraphics[width=.75\textwidth]{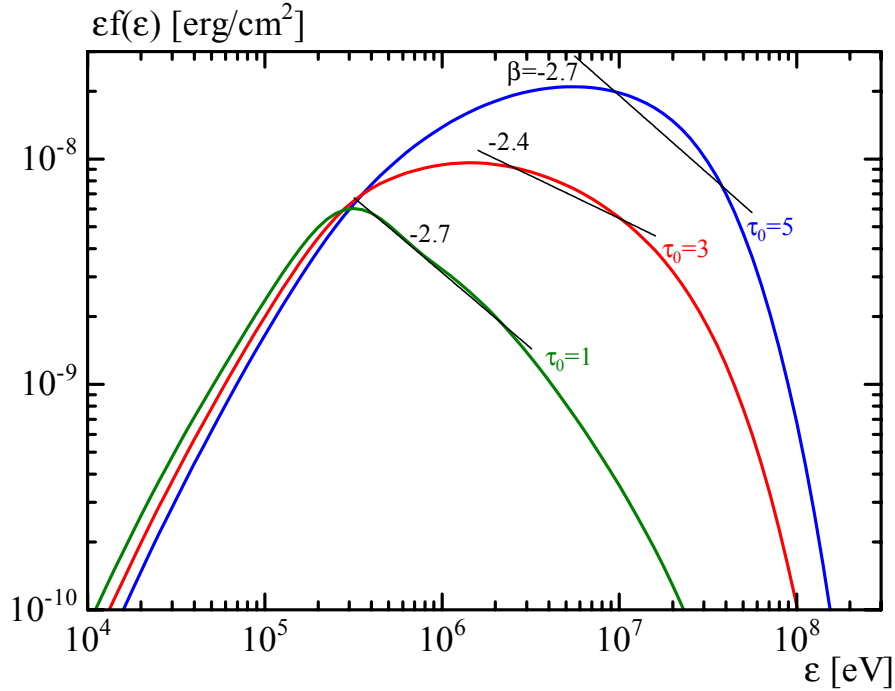}
\caption{Leptonic dissipative-photosphere models:
Time-integrated spectra for an observer
with different initial optical depths, $\tau_0=1$, 3, and 5.
The electron temperature evolution is assumed to be
$T'_{\rm e}=100$ keV $(R/R_0)^{-1}$.
Spectral fits with the power-law function
above $\varepsilon_{\rm p}$ are also shown.
\label{fig:3}}
\end{figure}

A different set of models, which we examine in more detail,
involves a relatively short duration of the energy-dissipation,
resulting in an electron temperature jump.
Figure \ref{fig:2} shows an example of the evolution of the
photon and electron distributions in the shell frame.
As shown in \cite{bel10}, for $\tau_0 \sim 1$,
$T_0 \sim 10$ keV is not enough to modify the Planck spectrum.
In this example we adopt $T_0=100$ keV, $\tau_0=5$,
and the index $s=1$.
In the initial stage, photons are efficiently up-scattered
by the hot electrons (see the lines for $R/R_0=1.0$--1.1).
At $R/R_0=1.1$ the photon spectrum may be well-fitted by
the Band function.
However, in the later stage, the peak energy shifts
to a higher energy.
As the electron temperature drops, the spectral curvature
becomes prominent and non-negligible.

Figure \ref{fig:3} shows the time-integrated spectra for an observer
(hereafter we assume the redshift $z=2$).
We try to fit the resultant spectra from the peak energy $\varepsilon_{\rm p}$
to $10 \varepsilon_{\rm p}$ by a power-law function.
The fitted results are shown in figure \ref{fig:3}.
For the low-energy region, the contribution
of the off-axis emission softens the spectra.
For the spectra in figure \ref{fig:3}, the low-energy photon indices are approximated
to $\alpha \simeq -0.5$.

Figure \ref{fig:3} illustrates, in summary form, the physical dependences
of leptonic dissipative-photosphere models.
The most critical parameter to determine the qualitative shape
of the spectrum is the optical depth at the radius where
the dissipation onsets.
The spectral shape in the case of $\tau_0=5$ does not match
a power-law for one order of magnitude above $\varepsilon_{\rm p}$.
On the other hand, the spectrum for $\tau_0=1$
is well fitted by a power-law function.
The spectrum for $\tau_0=3$ also shows a round feature,
but the typical error in the fluence in MeV band may be so large
that the curvature of such spectra are hard to detect.
However, for $\tau_0=3$ and $5$, the spectra below the peak
are also round and broad compared to the case for $\tau_0=1$.
To measure the broadness quantitatively,
we define the ``half-maximum energy'' $\varepsilon_{\rm h}$ that satisfies
\begin{eqnarray}
\varepsilon_{\rm h} f(\varepsilon_{\rm h})
=\frac{1}{2} \varepsilon_{\rm p} f(\varepsilon_{\rm p}),
\end{eqnarray}
where $\varepsilon_{\rm h} < \varepsilon_{\rm p}$.
A larger (smaller)
ratio $\varepsilon_{\rm h}/\varepsilon_{\rm p}$
means a sharper (broader) profile of the spectral peak.
For the Band function with $\alpha=-1.0$ and $-0.5$,
$\varepsilon_{\rm h}/\varepsilon_{\rm p}$
becomes $0.23$ and $0.35$, respectively.
In the cases for figure \ref{fig:3},
$\varepsilon_{\rm h}/\varepsilon_{\rm p}=0.1$,
0.13, and 0.39, for $\tau_0=5$, 3, and 1, respectively.
Since $\alpha \simeq -0.5$,
only the case of $\tau_0=1$ is consistent with the Band function.
This may imply that a large shift of $\varepsilon_{\rm p}$
due to the dissipation is not favorable.
For example, Axelsson et al. (2012) \cite{axe12}
claimed that the time-resolved spectra in GRB 110721A are
modeled with a combination of a Band function and a blackbody spectrum
(temperature $\sim 80$ keV).
The initial $\varepsilon_{\rm p}$ is as high as $\sim 15$ MeV so that they
suggested that the thermal photons are up-scattered to 15 MeV by electrons.
At least in our simulations, however, even when the scattering
is so efficient that $\varepsilon_{\rm p}$ shifts to a higher-energy,
a low-energy bump due to the initial thermal component does not appear
in the spectra.
The spectra in our results
seem to be single component with a smooth and broad shape.

\begin{figure}[htb!]
\centering
\includegraphics[width=.75\textwidth]{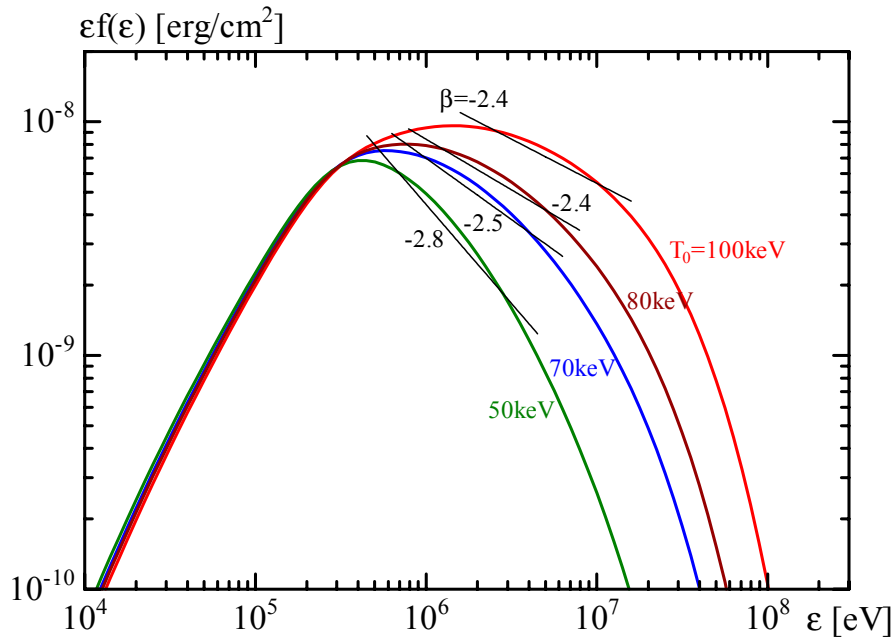}
\caption{Leptonic dissipative-photosphere models:
Initial electron temperature dependence
for time-integrated spectrum for an observer
in a model with the initial optical depth $\tau_0=3$.
The electron temperature evolution is assumed to be
$T'_{\rm e}=T_0 (R/R_0)^{-1}$.
Spectral fits with the power-law function
above $\varepsilon_{\rm p}$ are also shown.
\label{fig:4}}
\end{figure}

\begin{figure}[htb!]
\centering
\includegraphics[width=.75\textwidth]{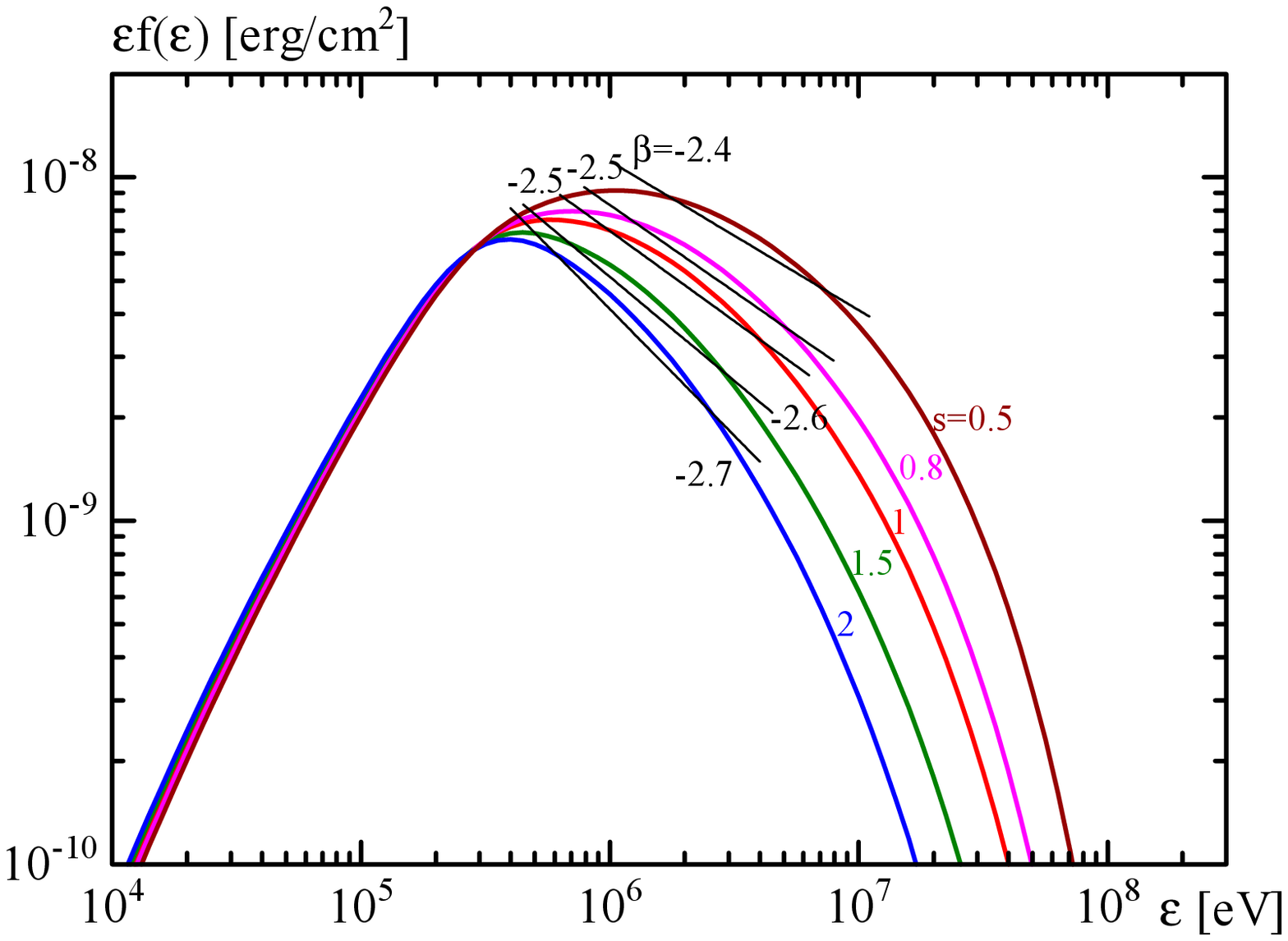}
\caption{Leptonic dissipative-photosphere models:
Electron-temperature evolution dependence
for time-integrated spectrum for an observer
in a model with the initial temperature $T_0=70$ keV
and optical depth $\tau_0=3$.
The electron temperature evolution is assumed to be
$T'_{\rm e}=T_0 (R/R_0)^{-s}$.
Spectral fits with the power-law function
above $\varepsilon_{\rm p}$ are also shown.
\label{fig:5}}
\end{figure}

\begin{figure}[htb!]
\centering
\includegraphics[width=.75\textwidth]{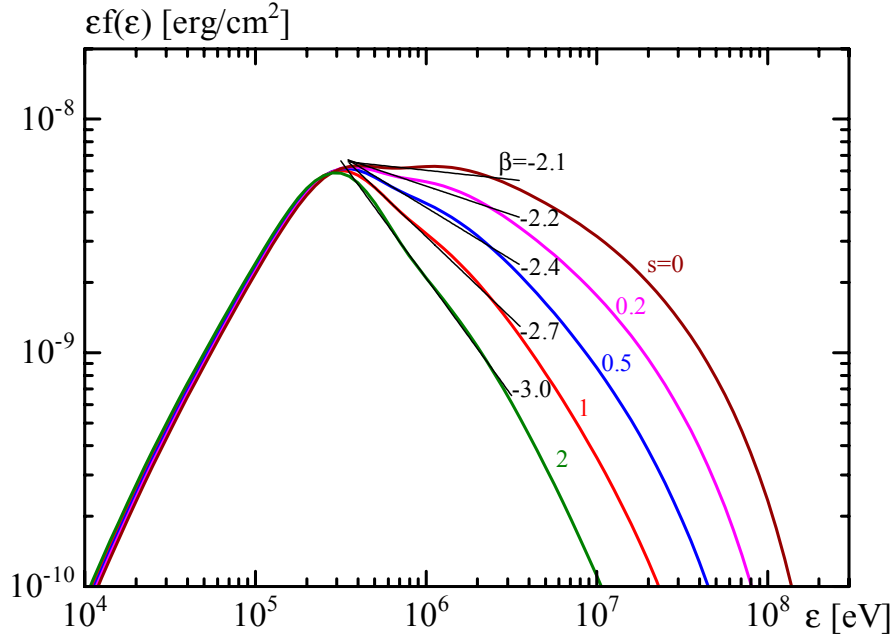}
\caption{Leptonic dissipative-photosphere models:
Electron-temperature evolution dependence
for time-integrated spectrum for an observer
in a model with the initial temperature $T_0=100$ keV
and optical depth $\tau_0=1$.
The electron temperature evolution is assumed to be
$T'_{\rm e}=T_0 (R/R_0)^{-s}$.
Spectral fits with the power-law function
above $\varepsilon_{\rm p}$ are also shown.
\label{fig:6}}
\end{figure}

We consider the possibility that the spectral curvature for a 
large $\tau_0$ can be determined by changes of the heating rate.
As shown in figures \ref{fig:4}--\ref{fig:6}, however,
given $\tau_0$, the degree of the deviation from a power-law
seems to not depend much on $T_0$ and the index $s$.
For $\tau_0=3$,
the spectra above $\varepsilon_{\rm p}$ are curved
irrespective of the heating-rate evolution.
Of course, for a larger $s$ or lower $T_0$,
the scattering efficiency becomes low enough to
make its spectrum as sharp as to be consistent with
the Band function. Namely,
the ratio $\varepsilon_{\rm h}/\varepsilon_{\rm p}$ can be
large enough for a small shift of $\varepsilon_{\rm p}$.
Figure \ref{fig:6} shows that the spectra for $\tau_0=1$
are well fitted by a power-law function for
a wide range of $s$.

\begin{figure}[htb!]
\centering
\includegraphics[width=.75\textwidth]{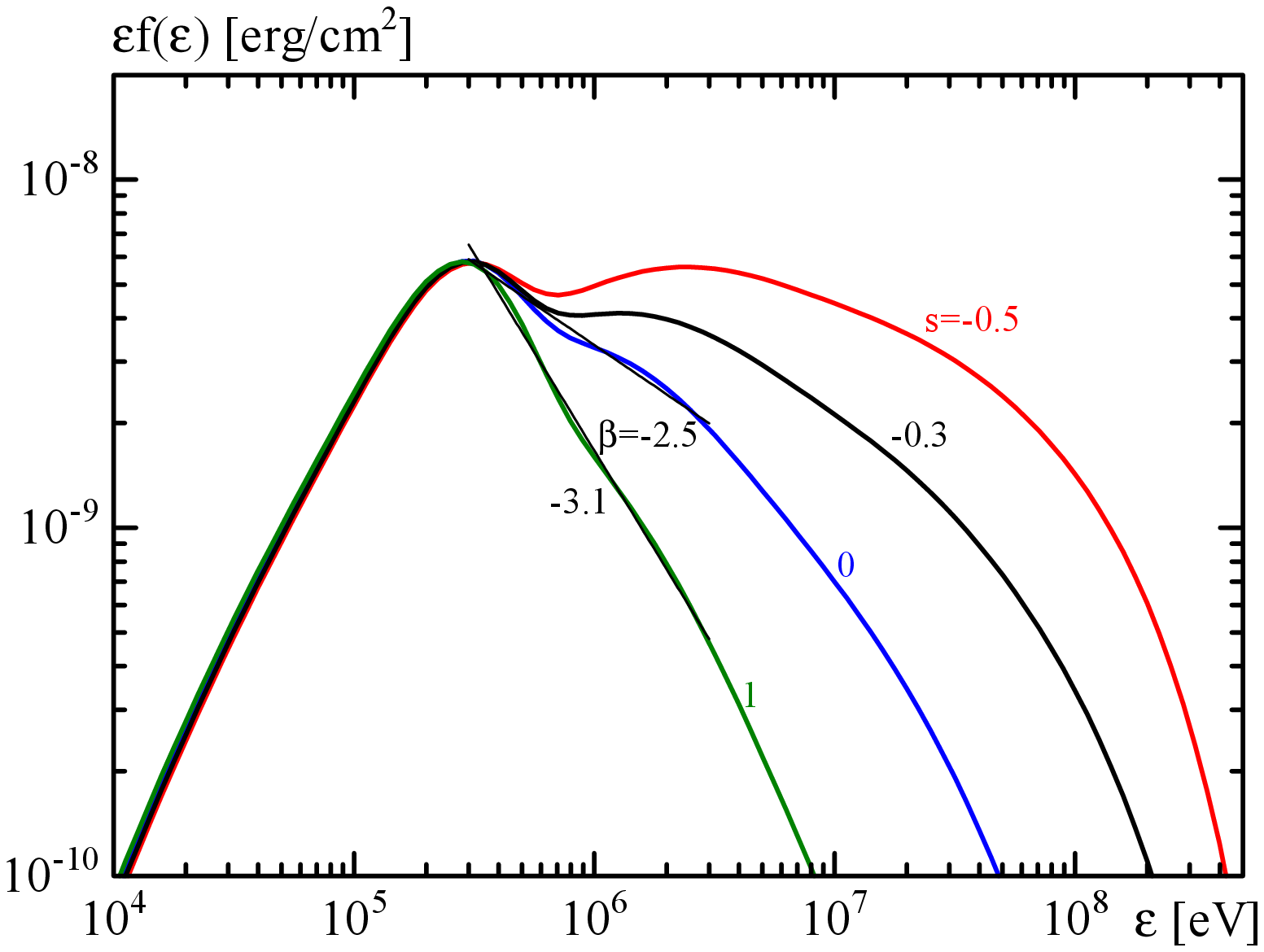}
\caption{Leptonic dissipative-photosphere models:
Electron-temperature evolution dependence
for time-integrated spectrum for an observer
in a model with the initial temperature $T_0=100$ keV
and optical depth $\tau_0=0.5$.
The electron temperature evolution is assumed to be
$T'_{\rm e}=\min(3 T_0, T_0 (R/R_0)^{-s})$.
Spectral fits with the power-law function
above $\varepsilon_{\rm p}$ are also shown.
\label{fig:7}}
\end{figure}

By analogy to the usual Compton $y$-parameter,
we can define the Compton amplification factor as
\begin{eqnarray}
y \equiv \frac{E_{\rm iso}-E_{\rm th}}{E_{\rm th}},
\label{eqy}
\end{eqnarray}
where $E_{\rm iso}$ and $E_{\rm th}$ are the final and initial
isotropic equivalent total energy in observed photons,
which can be compared with the analytical expectation,
\begin{eqnarray}
y_0=\frac{4 T_0}{m_{\rm e} c^2} \tau_0.
\end{eqnarray}
Of course, this conventional expression is for a static source.
As we will see below,
the actual value would be modified for an expanding plasma
so that we assume $y_0 \propto \tau_0$ even for $\tau_0>1$.
The results for figures \ref{fig:4}--\ref{fig:7}
are summarized in table \ref{tbl:1}.
Since the temperature and electron density drop with time,
$y$ obtained from our simulations tends to be lower than $y_0$.
When the electron temperature is maintained relatively high
($s \lesssim 0.5$),
the interaction between photons and electrons
becomes effectively longer than the dynamical timescale
at the initial radius.
In such cases, the resultant $y$ can exceed $y_0$.
A mildly-relativistic electron-temperature ($\sim 100$ keV)
also leads to a larger value of $y$ than $y_0$.

\begin{table}[htb!]
\begin{center}
\caption{The resultant Compton amplification factor (and $\varepsilon_{\rm p}$)
for Figs. \ref{fig:4}--\ref{fig:7} (from top to bottom).\label{tbl:1}}
\begin{tabular}{lccccc}
\hline \hline
$T_0$ (keV) & 50 & 70 & 80 & 100 &  \\ \hline
$\varepsilon_{\rm p}$ (keV) & 450 & 630 & 790 & 1600 & \\
$y$ ($y_0$) & 0.9 (1.2) & 1.6 (1.6) & 2.1 (1.9) & 3.2 (2.3) & \\
\hline \hline
$s$ & 2 & 1.5 & 1 & 0.8 & 0.5  \\ \hline
$\varepsilon_{\rm p}$ (keV) & 400 & 450 & 630 & 790 & 1100 \\
$y$ ($y_0$) & 0.9 (1.6) & 1.2 (1.6) & 1.6 (1.6) & 1.9 (1.6) & 2.7 (1.6) \\
\hline \hline
$s$ & 2 & 1 & 0.5 & 0.2 & 0  \\ \hline
$y$ ($y_0$) & 0.4 (0.8) & 0.7 (0.8) & 1.0 (0.8) & 1.5 (0.8) & 2.0 (0.8) \\
\hline \hline
$s$ & 1 & 0 & -0.3 & -0.5 &  \\ \hline
$y$ ($y_0$) & 0.3 (0.4) & 0.8 (0.4) & 1.4 (0.4) & 2.4 (0.4) &  \\
\hline
\end{tabular}
\end{center}
\end{table}

\begin{figure}[htb!]
\centering
\includegraphics[width=.75\textwidth]{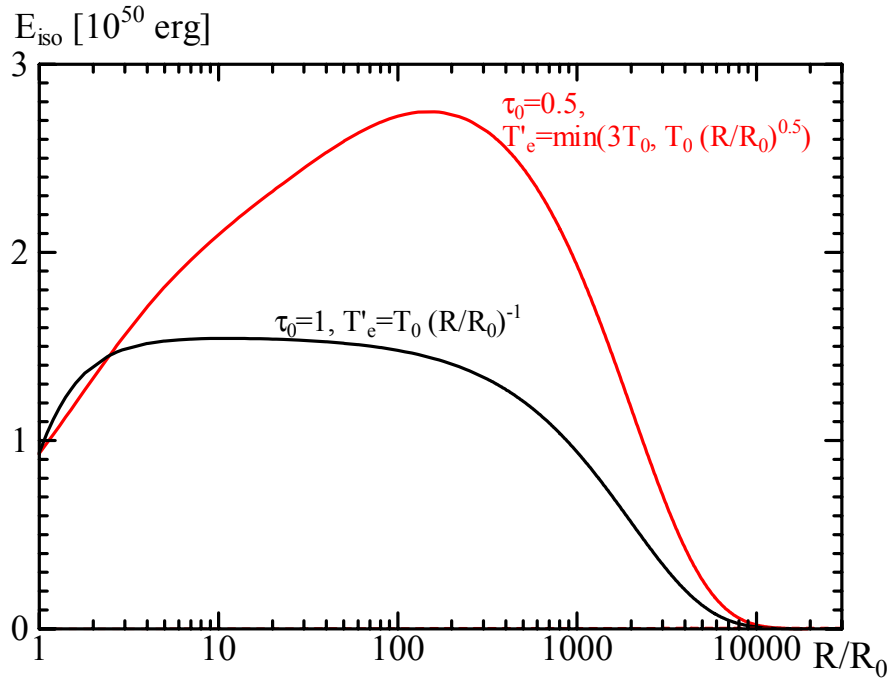}
\caption{Leptonic dissipative-photosphere models:
Photon-energy evolution within the shell
in models with the initial temperature $T_0=100$ keV.
\label{fig:8}}
\end{figure}

Even if $\tau_0$ is less than unity,
the high-energy spectra can be fitted with a power-law function
for $s>0$ (see figure \ref{fig:7}).
However, the low scattering efficiency leads to
a softer spectrum ($\beta<-2.5$) than the typical GRB spectrum
($-2.5 < \beta < -2.0$).
For $\tau_0=0.5$, to make the spectrum even harder,
we apply an increasing electron-temperature ($s<0$)
rather than decreasing-temperature.
We set an upper limit for the electron temperature as $3 T_0$
to maintain its non-relativistic temperature.
As shown in figure \ref{fig:7}, the obtained spectra become hard,
but they show a dip in their shape.
This is due to the late energy injection at large radii.
In spite of the low density at such radii, the mildly relativistic temperature
and long timescale lead to a sufficient energy-transfer to photons
as shown in figure \ref{fig:8},
which shows the evolution of the total photon-energy
within the shell.
Note that the electron-energy is negligible compared to
the photon energy before the photons escape.
Thus, the curves in figure \ref{fig:8} practically
show the energy-injection history due to the dissipation.
Such a gradual energy dissipation
at radii far from the photosphere may be achieved
only by the magnetic dissipation.

\section{Hadronic Dissipation Models: Spectral Shape}
\label{sec:had}

While we have discussed the role of the thermal electrons in the previous section,
non-thermal electrons may contribute to the modification of the Planck spectrum.
If a shock occurs below the photosphere, 
its structure is mediated by the radiation via Compton scattering \cite{bud10}.
The shock transition region for the radiation-dominated plasma
may be much thicker than the plasma skin depth and the gyroradius
of electrons \cite{lev12}.
Hence, a direct electron acceleration by the first order Fermi acceleration
may not work in the photosphere models.
Beloborodov (2010) \cite{bel10} proposed a two-fluid model, in which
nuclear collisions between the two fluids inject secondary particles.
If the relative Lorentz factor for the two fluids is a few,
the kinetic energy per proton belonging to the slower fluid
is a few GeV in the rest frame of the faster fluid.
The typical energy of the secondary pions produced
via $pp$ or $pn$-collisions is GeV.
A significant fraction of the pion energy quickly
converts to the non-thermal electron--positron pairs
via $\gamma \gamma$-absorption.
Such pairs may scatter the thermal photons to
higher energies and modify the spectrum as shown in \cite{bel10}.

In this section, we consider a similar situation to the case
in \cite{bel10}.
While the radiation transfer is calculated in steady flows
with pair injection in \cite{bel10},
our code is one-zone and time-dependent with hadronic processes.
One important difference in the computation is the hadronic process;
we numerically solve the $pn$-collisions and succeeding processes
for pions, muons, and electron--positron pairs such as
Compton/Thomson scattering, synchrotron, pair production, and
adiabatic cooling \cite[for details, see][]{asa12}.
Electron--positron pair annihilation is not included in our code,
but its contribution to the photon spectrum may not be prominent \cite{bel10}.
Since the injected pairs ($\varepsilon_{\rm e} \sim 0.5$ GeV)
lose most of their energy, cooling down to as low as $\sim 0.5$ MeV 
before annihilation, the energy fraction emitted as annihilation line 
emission is not large. We should also note that
there is so far no evidence for a 511 keV-line emission.

We promptly inject neutrons\footnote{We use the experimental $pp$ cross 
sections, although in \cite{bel10} a neutron flow was adopted.
However the difference between $pp$ and $pn$ cross sections is not 
essential for the hadronic cascade models.},
whose kinetic energy is almost monoenergetic at a nominal value of
$3$ GeV, at the initial radius $R_0$.
The number density of the target protons is assumed to be
the same as the electron density $n'_{\rm e}$ derived from $\tau_0$.
The cross section of $pp$-collision is typically
$\sim 0.05 \sigma_{\rm T}$ and the energy fraction
pions carry per collision (inelasticity) is 0.3
to the energy of parent protons.
A significant energy is carried by the target protons as well,
but we neglect the pion production from such recoiled protons/neutrons
for simplicity.
The number fraction of neutral pions, which efficiently inject pairs
via electromagnetic cascade, is roughly 1/3 to the
total number of pions.
Hence, the photon energy extracted from the protons
may be approximated as $5 \times 10^{-3} \tau_0 E_{\rm p}$,
where $E_{\rm p}$ is the total energy of the injected protons.
Here, we assume that the role of the background electrons is sub-dominant
compared to that of protons.
So we adopt a low temperature for the background electrons
as $T_0=10$ keV and $s=1$ [see equation (\ref{eqTev})],
which is close to the model in \cite{bel10}.
The magnetic field is assumed to evolve as
$B'=B_0 (R/R_0)^{-1}$, where $B_0=1.4 \times 10^5$ G.
This implies the initial energy density ratio
of the magnetic field to photons is $U'_B/U'_\gamma=10^{-4}$
for $L_{\rm th}=10^{52} \mbox{erg}~\mbox{s}^{-1}$.

\begin{figure}[htb!]
\centering
\includegraphics[width=.75\textwidth]{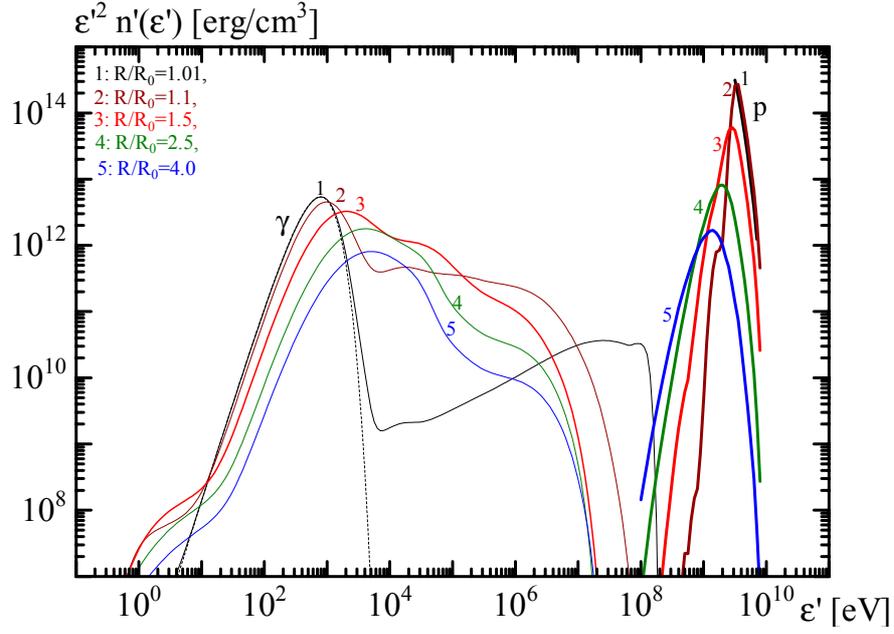}
\caption{Hadronic dissipative-photosphere model:
Evolution of the energy distribution of photons (thin)
and protons (thick) in the shell rest frame for a model
with $\tau_0=30$ and $W=10^3 R_0/\Gamma^2$.
\label{fig:9}}
\end{figure}

Figure \ref{fig:9} shows an example of the evolution
of the photon and proton energy distributions in the shell rest frame.
Here, we adopt $W=W'/\Gamma=10^3 R_0/\Gamma^2$ as in the previous section.
The injected proton energy is $E_{\rm p}=2 \times 10^{51}$ erg,
which implies the initial ratios $E_{\rm p}/E_{\rm th}
=U'_{\rm p}/U'_\gamma=21.6$.
We adopt $\tau_0=30$, which determines the densities of
the background electrons and protons.
Neglecting the thermal energy,
the total energy of the background protons
is still larger than $E_{\rm p}$, roughly $E_{\rm p,bg}/E_{\rm th} \simeq 55$.
From the above parameters,
we can expect $y \sim 3$ [defined in eq. (\ref{eqy})].
As pairs are injected, the thermal photons are up-scattered
as far as $10^8$ eV.
Those photons are absorbed soon and
the spectrum evolves to a shape similar to the Band function
at $R/R_0=1.5$.
In the later stage, however, the low-temperature electrons
down-scatter high-energy photons.
The peak energy gradually shifts towards higher energies as the scattering
process by the thermal electrons proceeds.
As a result, the spectrum has a broad peak around 10 keV in the shell frame.
We can see the effects of the adiabatic cooling and volume expansion
in the evolution of the proton density distribution.

\begin{figure}[htb!]
\centering
\includegraphics[width=.75\textwidth]{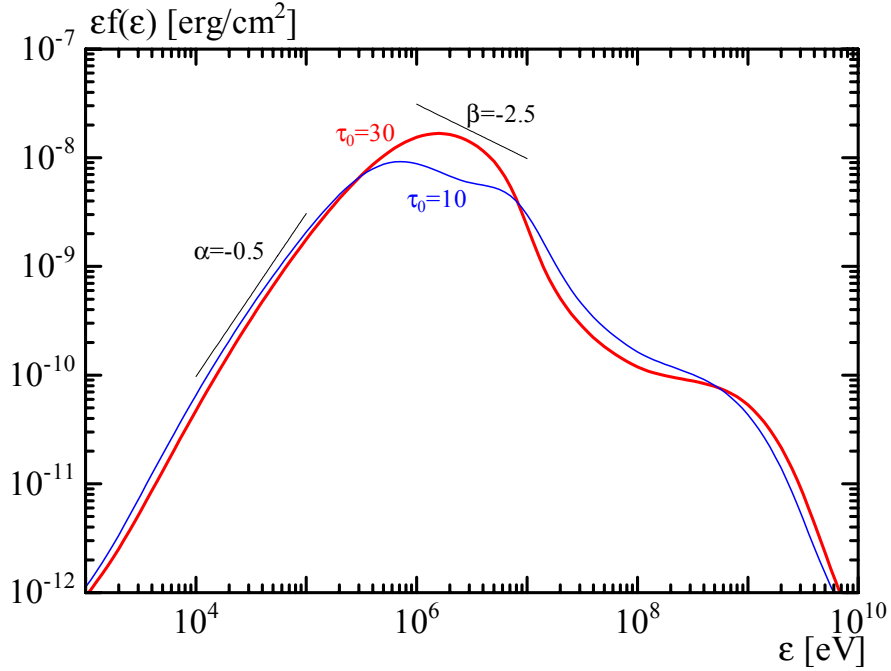}
\caption{Hadronic dissipative-photosphere model:
Observer frame time-integrated spectra for 
models
with $W=10^3 R_0/\Gamma^2$
and different initial optical depths, $\tau_0=30$ and 10.
\label{fig:10}}
\end{figure}

The time-integrated spectra for an observer is shown in
figure \ref{fig:10}. The spectrum for $\tau_0=30$ has a broad peak
as expected from figure \ref{fig:9}.
A lower $\tau_0$ leads to a less efficient interaction with the background electrons.
In this case, the thermal electrons do not modify the spectrum very much
so that the spectrum for $\tau_0=10$ shows a power-law-like spectrum
for one order of magnitude above $\varepsilon_{\rm p}$.
The lower $\tau_0$ also leads to a lower efficiency for $pp$-collision.
For $\tau_0=10$, the initial ratio $E_{\rm p,bg}/E_{\rm th} \simeq 18$
implies $E_{\rm p}>E_{\rm p,bg}$
(note that we have fixed the ratio $E_{\rm p}/E_{\rm th}=21.6$).
The amplification factor [see eq. (\ref{eqy})] are
4.0 and 2.4 for $\tau_0=30$ and 10, respectively.

\begin{figure}[htb!]
\centering
\includegraphics[width=.75\textwidth]{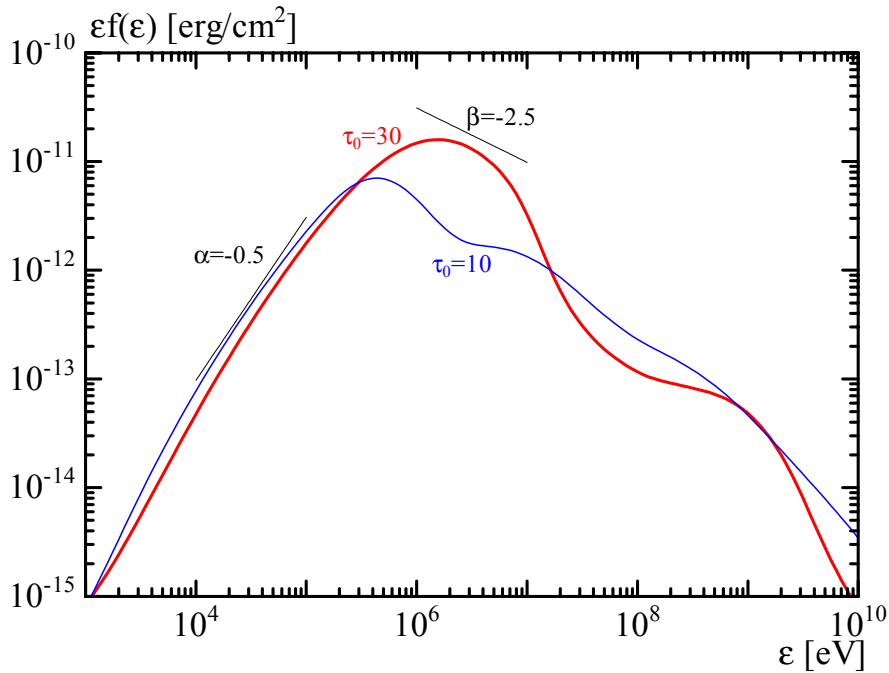}
\caption{Hadronic dissipative-photosphere model:
Same as Fig. \ref{fig:10} but for
thin shell models ($W=R_0/\Gamma^2$).
\label{fig:11}}
\end{figure}

The fluences in the GeV range in figure \ref{fig:10}
are much lower than the power-law extrapolations
from $\varepsilon_{\rm p}$.
In order to have a relatively higher GeV fluence in the hadronic
dissipative-photosphere model,
we test a thin shell model of $W=R_0/\Gamma^2$.
We preserve the luminosity $L_{\rm th}$ and the ratio $E_{\rm p}/E_{\rm th}$.
Thus, the smaller volume leads to $E_{\rm p}=2 \times 10^{48}$ erg.
If we consider a case similar to the internal shock, the 
interaction of the two colliding flows (or the $pn$ decoupling) starts
at a certain radius $R_0$.
Then, the causal width of the flow region affected by the other flow
in the dynamical timescale is $R_0/\Gamma^2$.
In such cases, the thin shell model seems more plausible.
In the thin-shell model,
photons quickly escape with a timescale of $\tau_0 R_0/\Gamma$
in the shell frame.
We neglect the interaction of escaped photons with the background flows.
Figure \ref{fig:11} shows the time-integrated spectrum for
this model with $\tau_0=30$ and 10.
The spectrum for $\tau_0=10$ can be fitted by a power-law with $\beta=-2.5$
within actual observational-errors, while $\tau_0=30$ results
in a broad peak without relatively significant GeV-flux again.
In conclusion, to have a power-law spectrum in a wide energy range,
the ideal case would be $\tau_0 \sim 10$ with a thin shell.
An even smaller $\tau_0$ is not favorable to induce $pp$-collision.

\begin{figure}[htb!]
\centering
\includegraphics[width=.75\textwidth]{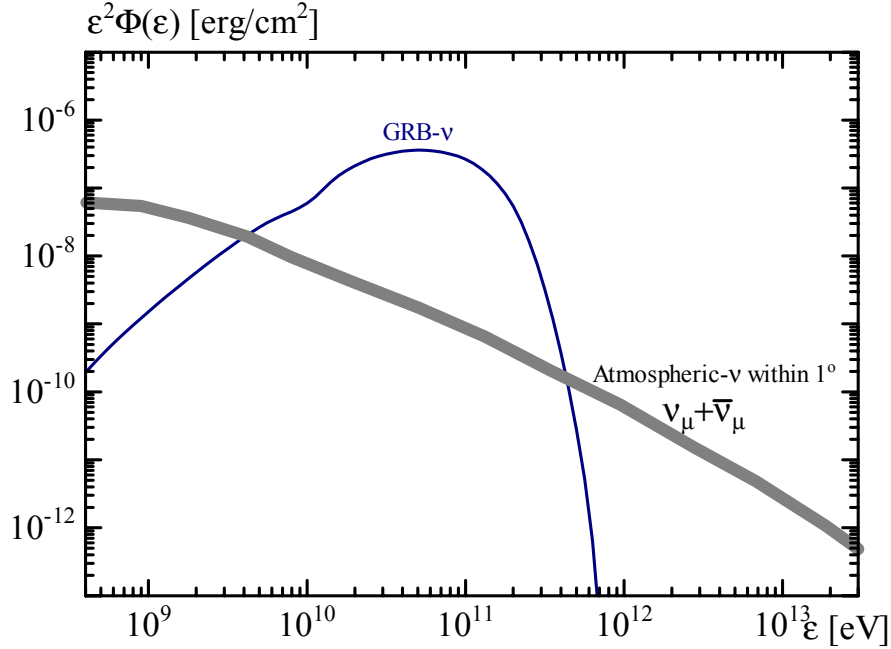}
\caption{Hadronic dissipative-photosphere model:
Neutrino spectrum from a thin-shell GRB with
$\tau_0=10$ and $N=10^5$ pulses ($E_{\gamma,{\rm iso}} \simeq 4.4 \times 10^{52}$ erg)
at $z=2$. The atmospheric neutrino fluence within an angle of
$1^\circ$ with 10-s integral is also plotted.
\label{fig:12}}
\end{figure}

Our numerical code simulates one pulse emitted from a shell,
including both the electromagnetic output as well as 
the neutrino pulse spectrum resulting from the $pn$
collision induced cascade.
In figure \ref{fig:12}, we plot {the total neutrino spectrum,
without distinguishing the neutrino species, for the thin-shell model 
with $\tau_0=10$. This corresponds to the model of Fig. \ref{fig:11}, 
which reproduces the observed photon spectrum.
To reproduce the typical total energy release from a burst,
we have assumed that $N=10^5$ identical pulses are emitted
within 10 s.
This implies an isotropic photon energy including the cascade effects
of $4.4 \times 10^{52}$ erg.
We also plot the background neutrino spectrum
(atmospheric neutrinos) in a 10 s window within a cone of an angle of $1^\circ$.
If this spectrum represents the typical neutrino spectrum from a burst
and the uncertainties in the neutrino directions are $1^\circ$,
we can expect a neutrino flux which is 10--100 times brighter than the 
background flux above 10 GeV.
However, the expected event rate with a detector such as Super-Kamiokande (SK)
is very low.
A time- and direction-coincidence analysis ($\pm 1000$ s and $\leq 15^\circ$)
using 1454 GRBs data with SK
provides an upper limit $\sim 1~\mbox{cm}^{-2}$ {per burst}
for $\nu_\mu$ and $\bar{\nu}_\mu$
at 40 GeV \cite{fuk02},
while our result is
$\varepsilon \Phi(\varepsilon) \simeq 5.4 \times 10^{-6}~\mbox{cm}^{-2}$ at 40 GeV.
The vacuum neutrino oscillations would lead to one-third of the above flux
appearing as a $\nu_\mu$ plus $\bar{\nu}_\mu$ flavor flux.
Although the new subarray of the IceCube, DeepCore,
is designed to lower the neutrino energy threshold,
the effective area for $30$--$100$ GeV is not much larger than
$10~\mbox{cm}^{2}$ \cite{abb12}.
Even if we lower the redshift to $z=0.1$,
the expected detection-rate is $\sim 3 \times 10^{-3}$
$\mu$-neutrinos per burst.

\section{Lightcurves}
\label{sec:LC}

For the thick shell case ($W=10^3 R_0/\Gamma^2$),
our one-zone homogeneous-shell approximation is not optimal
for computing lightcurves in the observer frame, because it is 
difficult to reconcile the homogeneity of the shell
with the relativity of simultaneity.
However, our tentative results for the leptonic model
(which we omit to show here, for the above reason)
agree with the simple expectation:
a very sharp rise due to the short timescale
at the photosphere $(1+z) R_0/\Gamma^2 \ll$ ms,
and decaying long tail with FWHM of $23$ ms,
which is close to $(1+z) W/c \simeq 28$ ms.
The very sharp rising time is an artifact of the homogeneous-shell
approximation.
In other words, the discrete photon-distribution
due to the sharp edge of the shell surfaces causes
the sharp rise.
If the outflow gradually evolves with a timescale of $\sim 0.1$ s,
the rise timescale may be similarly extended.
In photosphere models, however, we may expect very short timescale ($\ll$ ms)
to be common properties of lightcurves.
The lightcurves for the hadronic thick-shell models are almost the same
as for the thick-shell leptonic models.

\begin{figure}[htb!]
\centering
\includegraphics[width=.75\textwidth]{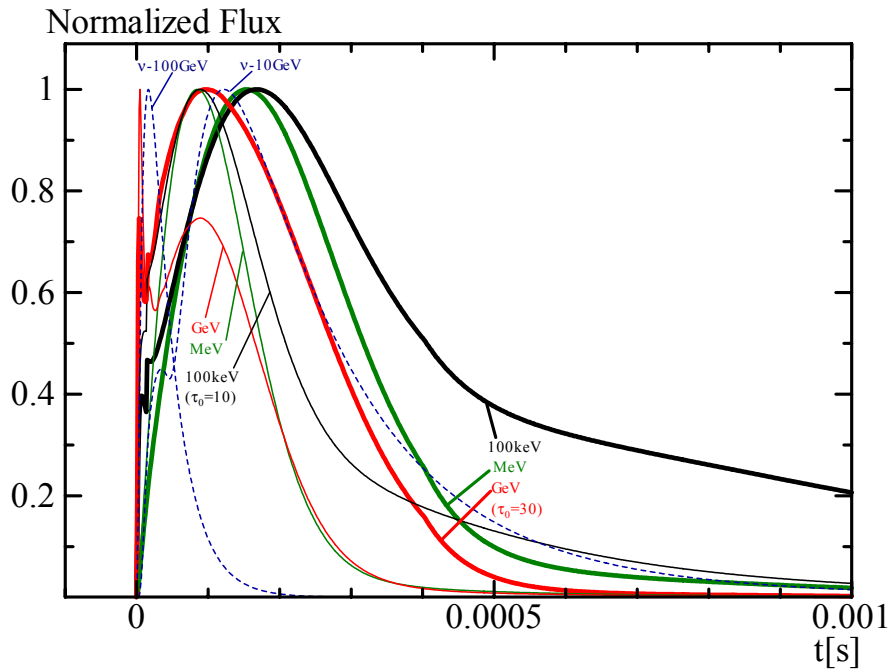}
\caption{Single-pulse lightcurves for hadronic dissipation models
with different initial optical depths, $\tau_0=30$ (thick solid)
and 10 (thin solid).
The shell width is assumed to be $W=R_0/\Gamma^2$.
Neutrino lightcurves (thin dashed) are for $\tau_0=10$.
\label{fig:13}}
\end{figure}

Figure \ref{fig:13} shows the lightcurve for the thin-shell hadronic models.
We expect the pulse timescale to be
\begin{eqnarray}
\Delta t_{\rm obs}=(1+z) \tau_0 \frac{R_0}{c \Gamma^2}
\simeq 2.8 \times 10^{-4} \left( \frac{\tau_0}{10} \right)~\mbox{s},
\end{eqnarray}
which roughly agrees with the numerical results in figure \ref{fig:13}.
The constant shell thickness $R_0/\Gamma^2$ is thinner
than the causal length scale, $\tau_0 R_0/\Gamma^2$,
at the photosphere. This thin shell results in efficient photon escape,
which slightly shortens the pulse timescale.
The initial sharp spikes seen in the GeV lightcurves
are the emission before the photon density grows enough
to absorb GeV photons via $\gamma \gamma$-process.
As the photon density grows in the dynamical timescale
$R_0/c \Gamma^2 \sim \Delta t_{\rm obs}/\tau_0$,
the GeV flux starts to decay.
The hadronic cascade continues longer than this timescale.
The deposited GeV photons in the shell are finally
released with the timescale of $\Delta t_{\rm obs}$.
For $\tau_0=30$, the GeV emission ceases earlier than MeV emission.
The down-scattering effect explained in \S \ref{sec:had}
gradually softens the spectrum, which causes the early termination of GeV emission.
The relatively long tail in the 100 keV-lightcurve
is due to the emissions from off-axis regions.
The spectrum above $\varepsilon'_{\rm p}$
in the shell frame is so soft that
the effect of the off-axis emissions is prominent only for
100 keV-lightcurve.
For $\tau_0=10$, the softening effect is only slight.
In this case, the peak times for MeV and GeV lightcurves
coincide with each other.

We also plot the neutrino lightcurves.
As shown in figure \ref{fig:12}, the neutrino peak energy
is roughly 100 GeV.
The 100 GeV neutrino lightcurve has a shorter timescale
than gamma-ray lightcurve owing to the lack of diffusive scattering in the shell,
but longer than the initial dynamical timescale
$(1+z) R_0/c \Gamma^2 \simeq 2.8 \times 10^{-5}$ s.
This timescale may be determined by the duration
of the $pp$-collisions in the highest energy range of protons.
The onset of the 10 GeV neutrino emission is 
delayed relative to the gamma-ray lightcurves,
and its timescale is longer.
The lower energy neutrinos are produced by protons which have suffered
adiabatic/dissipative cooling subsequent to injection, the  cooling timescale
leading to the delayed and longer lasting  neutrino emission in
the 10 GeV range plotted.

If we adopt the thin shell for the leptonic model as well,
the lightcurve shape is simple and similar to that
for the internal shock model in \cite{asa11} except for
the very short pulse timescale.
For the models with $\tau_0=1$,
a significant delay between MeV and 0.1-1 GeV lightcurves
is not seen in our one-zone model, hence as mentioned
such delays may require multi-zone models.
Note that the GeV flux is too dim to observe in our leptonic model.

Note that while in figure \ref{fig:13} we show the lightcurves for a single pulse, 
the neutrino spectral flux of figure \ref{fig:12} is based on the $N=10^5$ pulses.
The individual pulses of short timescale ($\ll$ ms) assumed here in the thin shell models
are likely to be hard to detect, with the typical photon/neutrino statistics 
expected or obtained with present instruments.
On the other hand, the typical ``observed'' pulse timescales are
$\sim 0.1$--$1$ s \cite{nor96}, which could be the result of pulses lumping
together into $\sim 10^3-10^4$ clusters of pulses, each of $0.1$--$1$ s duration.
In this case, the $N=10^5$ pulses assumed in figure \ref{fig:12}
are divided into $10$--$100$ clusters (``pulses observationally defined'')
with an individual $0.1$--$1$ s timescale within the total 10 s-duration.
This pulse clustering may be a feature intrinsic to the long-term variability
of the central engine activity. It is thus likely that this intrinsic
variability would wash out the sharp features of individual pulses,
depending on the interval between pulses.

\section{Summary and Discussion}
\label{sec:summary}

Using our time-dependent code, we have simulated the GRB photospheric 
emission for models with both a gradual and a sudden onset of the energy 
dissipation below and around the photosphere.

For the models with gradual energy dissipation, e.g. caused by gradual
magnetic reconnection, our numerical simulations show a time evolution
of the photon energy distribution in general agreement with the
previous results of \cite{gia06,gia12}, which lead to generic
Band-like spectral forms in general agreement with observations.

Another class of models studied here involves a sudden onset of 
dissipation, starting at some depth below the photosphere, 
involving either leptonic or hadronic  mechanisms.

For the leptonic models discussed here, the observed Band spectra are
best reproduced when the up-scattering  of seed thermal photons by 
dissipation-heated thermal electrons starts around $\tau_0\sim 1$, 
irrespective of the heating history.
For larger initial optical depths, a curved spectrum tends to form, which
deviates from a power-law shape. However, due to observational uncertainties 
in the 1--10 MeV observed spectra, cases with  $\tau_0 \simeq 3$ are still
acceptable. For a given electron temperature, smaller optical depths reduce 
the energy injection rate, but in such cases the high-energy 
spectral slope becomes softer than the observations, unless the energy 
injection is gradual and extended in time. Such long energy injections 
(``slow heating") may be characteristic of magnetic dissipation models.
Even in such models, deviation from a power-law spectrum  is expected
(spectral dips,  etc. - see \S \ref{sec:lep} ).
Since  the universality of the Band function as a simple power-law extending 
far above $\varepsilon_{\rm p}$ is not certain, improvements in the photon 
statistics in this energy range would be helpful in testing for evidence of 
such leptonic dissipative photospheres.

A common expectation for our leptonic sudden heating models
is a significant deviation from the Band spectrum above 10 MeV,
as long as the thermal electrons are non-relativistic. This will be also 
tested through continued observations with {\it Fermi} \cite{ack12}.
The inferred ideal optical depth of $\tau_0=1$ may imply that the onset of 
the dissipation is controlled by the optical depth. This is reminiscent
of the reconnection switch model of \cite{mck12}.

A significant shift of the peak energy due to a large
energy injection through dissipation is not likely,
because, in such cases, the spectrum becomes broader than
the typical Band function. Also, in such cases the obtained spectrum is 
not a superposition of a thermal and Band spectra. Thus, if the low-energy 
bump due to the thermal component is present as reported in \cite{gui11} 
and \cite{axe12}, two different emission regions may be required.

In the hadronic sudden heating models based on $pn$-collision cascades, 
the ideal optical depth for initiating the process is $\tau_0 \sim 10$,
in order to extract enough energy from hadrons and reproduce the Band 
spectrum \cite{bel10}.
The densities are higher than in the leptonic model, which tends to
produce a spectral bump due to the background thermal electrons.
This requires a careful choice of parameters in order to avoid 
an excessive effect of the thermal electrons.  This model naturally
predicts neutrino emission in the 10-100 GeV range. If such 
photosphere models also lead to proton acceleration, even higher 
energy neutrinos would also be produced \cite{gao12}.
The difference in the neutrino spectrum would be an important clue
for investigating the emission and particle acceleration mechanisms.
However, the detection of the predicted neutrino flux levels with 
present detectors is difficult for the $pn$-collision models, 
as also shown in \cite{mur13} \cite[see also][]{bar13}.

In both types of photospheric models discussed here, the GeV photon onset 
is almost the same or earlier than the MeV onset, unlike the delayed GeV 
emission phenomenon reported in several {\it Fermi}-LAT GRBs. Thus, in
photosphere models the reason for the delayed onset may not be an 
intrinsic property of the MeV emission regions; instead, more complex,
e.g. two-zone models such as \cite{tom11} \cite[see also][]{asa11} 
may be required, at least for {\it Fermi}-LAT GRBs.

The low-energy spectral index $\alpha \sim -0.5$ in our calculations 
is softer than the Planck spectrum, caused by the softer contributions
from the off-axis emission (remembering that here we adopted 
$\theta_{\rm j}=10/\Gamma$).  In order to further soften the spectrum 
to values closer to the typical observed index $-1$, some kind of 
inhomogeneity depending on polar angle may be needed, as shown 
in \cite{lun13}. On the other hand, as seen in some time-resolved spectra 
such as in GRB 090902B \cite{902B}, a very hard index $\alpha > -0.5$
might  imply a less significant contribution of the off-axis emissions.
This would mean $\theta_{\rm j} \leq 1/\Gamma$, which may result in an very 
early jet break in the afterglow lightcurves. However, the X-ray and optical 
afterglow in GRB 090902B does not show a jet break during the first 6 days 
\cite{pan10}, which suggests a large jet opening-angle $\theta_{\rm j}>0.11$.
This is contradiction to the above simple one-zone model, indicating a 
limitation on such one-zone calculations. Other limitations are indicated by
the observation of a shallow decay phase and chromatic breaks in a number
of GRB afterglows with {\it Swift}. To explain such features may
require multi-zone models of the emission region \cite[e,g.][]{rac08},
similarly to the two-zone 
models for the prompt emission \cite[e.g.][]{tom11,asa11}. The increased 
complexity of such geometrical structure may soften the above contradiction 
between harder spectra and late jet breaks.

The ideal parameters for sudden heating models to
reproduce the Band spectrum cluster within relatively
narrow  ranges  around  $\tau_0 \sim 1$ with $T_0=100$--$300$ keV for
the leptonic case, and $\tau_0 \sim 10$ for the hadronic case.
This may be interpreted as a requirement for the dissipation mechanism, 
providing a test for selecting between competing possibilities,
such as the gradual magnetic reconnection \cite{gia06,gia07,mck12},
hydrodynamical turbulence \cite{zha09,miz11,ino11},
plasma instability \cite{iok07}, $pn$($pp$)-collision \cite{bel10},
or usual internal shocks \cite[etc.]{pee06}. At this point, the study of
these mechanisms is not specific enough to subject them to such tests.
Future observational and theoretical (numerical) studies of the various 
dissipation mechanisms and their specific spectral predictions will be
needed in order to pinpoint which, if any, mechanisms are robustly
capable of reproducing the spectral shape within the context of a 
specific dissipative photosphere model.

Finally, we briefly comment on the radiation efficiency of the photospheric models.
Our method in this paper does not treat the dissipation mechanism itself,
which determines what fraction
of the jet energy in dissipated into gamma-rays.
The models whose spectrum is consistent with the Band function
require the dissipation energy of $\sim 1$--$2$ times the initial
thermal photon energy as shown in Table \ref{tbl:1}.
In the standard fireball model, the fraction of
the thermal photons to the total in luminosity at the photosphere is
\begin{eqnarray}
\frac{L_{\rm th}}{L_{\rm tot}}=
\left( \frac{4 \pi \Gamma^4 R_0 m_{\rm p} c^3}{L_{\rm tot} \sigma_{\rm T}}
\right)^{\frac{2}{3}} \sim 0.1 \left( \frac{\Gamma}{600} \right)^{\frac{8}{3}}
\left( \frac{R_0}{100~\mbox{km}} \right)^{\frac{2}{3}}
\left( \frac{L_{\rm tot}}{10^{54}\mbox{erg}~\mbox{s}^{-1}} \right)^{-\frac{2}{3}},
\end{eqnarray}
where $R_0$ is the initial size of the fireball.
The fraction largely depends on $\Gamma$. If $\Gamma \gtrsim 600$,
it would be a few tens of percent.
In such cases, the leptonic model requires a large fraction of the bulk kinetic energy
to be dissipated in order to add extra photons of $\sim L_{\rm th}$, 
and the gamma-ray emission efficiency should be quite high, $\gtrsim 0.4$.
On the other hand, the hadronic model requires at least
$E_{\rm tot}/E_{\rm th} \gtrsim 30$ to cause significant $pp$-collisions.
Hence, the radiation efficiency may be $\lesssim 0.1$.

\begin{acknowledgments}
We thank the anonymous referee for valuable comments.
This study is partially supported by Grants-in-Aid for Scientific Research
No.22740117 and No.25400227 from the Ministry of Education,
Culture, Sports, Science and Technology (MEXT) of Japan,
and NASA NNX13AH50G.
We also thank K. Ioka for valuable comments.
\end{acknowledgments}

\end{document}